\documentclass{article}
\usepackage[papersize={210mm,297mm}, total={160mm,250mm}]{geometry}

\usepackage[square,numbers]{natbib}
\usepackage[utf8]{inputenc} 
\usepackage[T1]{fontenc} 
\usepackage{hyperref} 
\usepackage{url} 
\usepackage{booktabs} 
\usepackage{amsfonts} 
\usepackage{nicefrac} 
\usepackage{microtype} 
\usepackage{lipsum}
\usepackage{fancyhdr} 
\usepackage{graphicx} 
\graphicspath{{media/}} 
\usepackage{xcolor}
\usepackage[para]{threeparttable}

\usepackage{amsmath}
\usepackage{nccmath}
\usepackage{tabularx}
\usepackage{multirow}
\usepackage{amssymb}
\usepackage{subfig}

\usepackage{soul}
\usepackage{color}
\sethlcolor{white}
\usepackage{listings}

\graphicspath{ {images/} }
\title{ Chat2VIS: Generating Data Visualisations via Natural Language using ChatGPT, Codex and GPT-3 Large Language Models

}

\author{
Paula Maddigan\footnote{\href{mailto:t.susnjak@massey.ac.nz}{P. Maddigan Email: Paula.Maddigan.1@uni.massey.ac.nz}}  and 
  Teo Susnjak\footnote{\href{mailto:t.susnjak@massey.ac.nz}{T. Susnjak Email: t.susnjak@massey.ac.nz}} \\
  School of Mathematical and Computational Sciences \\
  Massey University \\
  Auckland, New Zealand \\
}

\begin{document}
\maketitle

\begin{abstract}
The field of data visualisation has long aimed to devise solutions for generating visualisations directly from natural language text. Research in Natural Language Interfaces (NLIs) has contributed towards the development of such techniques. However, the implementation of workable NLIs has always been challenging due to the inherent ambiguity of natural language, as well as in consequence of unclear and poorly written user queries which pose problems for existing language models in discerning user intent. Instead of pursuing the usual path of developing new iterations of language models, this study uniquely proposes leveraging the advancements in pre-trained large language models (LLMs) such as ChatGPT and GPT-3 to convert free-form natural language directly into code for appropriate visualisations. This paper presents a novel system, Chat2VIS, which takes advantage of the capabilities of LLMs and demonstrates how, with effective prompt engineering, the complex problem of language understanding can be solved more efficiently, resulting in simpler and more accurate end-to-end solutions than prior approaches. Chat2VIS shows that LLMs together with the proposed prompts offer a reliable approach to rendering visualisations from natural language queries, even when queries are highly misspecified and underspecified. This solution also presents a significant reduction in costs for the development of NLI systems, while attaining greater visualisation inference abilities compared to traditional NLP approaches that use hand-crafted grammar rules and tailored models. This study also presents how LLM prompts can be constructed in a way that preserves data security and privacy while being generalisable to different datasets. This work compares the performance of GPT-3, Codex and ChatGPT across a number of case studies and contrasts the performances with prior studies.

\end{abstract}



\maketitle

\section{Introduction}\label{}

The ability to generate visualisations based on natural language (NL) text has long been a desirable goal in the field of data visualisation. 
Research into Natural Language Interfaces (NLIs) for visualisation has emerged as the primary field that has recently spearheaded the advancements in this area \cite{narechania2020nl4dv,shen2021towards}. These interfaces allow users to generate visualisations in response to NL queries or prompts that are free from programming and technical constructs, thus providing a flexible and intuitive way to interact with data. The ultimate aim is to devise systems enabling users to express queries like \textit{"Show me the sales trend?"} which are correctly understood and depicted by automatically discerning the correct chart type. 

The data visualisation paradigm can be difficult to learn for users \cite{wang2022towards} who must translate their analysis intentions into tool-specific operations which may take the form of point-and-click applications or code in various programming languages. Therefore, NLIs can improve the usability of visualisation tools \cite{luo2021natural} by making them more convenient and novice-friendly as well as effective, and ultimately, inclusive for a broader range of users. As such, approaches like NLIs have the potential to make data and insights more accessible to a wider audience and lower the barrier \cite{wang2022towards} by allowing users to express their queries and analysis intentions in a form that is most natural to them \cite{wang2021survey}.

The process of translating NL inputs into visualisations (NL2VIS) involves several non-trivial tasks. Generally, the input query is first parsed and modelled, then the required data attributes are identified, and the low-level analytic tasks expressed within the query are discerned. These low-level tasks, such as filtering, correlation, and trend analysis, must then be translated into code to be executed. Finally, the input query is analysed and matched with the most appropriate visualisation, and then code is invoked to render the data. Each component in the pipeline is error-prone.

The implementation of NL2VIS is a particularly challenging task due to the inherent characteristics of NL, such as ambiguity and the underspecification of requirements in the prompts, as well as unavoidable typographical errors. These characteristics of NL make it difficult to accurately interpret the user's intent and generate appropriate visualisations with the existing technologies and approaches \cite{song2022rgvisnet}. Despite these challenges, the popularity of NLIs for data visualisation has continued to grow, driven by the demand for data analytics and the increasing need for flexible and intuitive ways of interacting with data.

The performance of NLIs for visualisations is largely dependent on Natural Language Processing (NLP) models \cite{wang2021survey} and their robustness in understanding NL. A recent comprehensive survey \cite{wang2021survey}  of the use of NLIs for visualisations noted that while most existing approaches utilise hand-crafted grammar rules that require proficiency with typical NLP toolkits, more complex large language models (LLMs) like GPT-3 which are capable of achieving human-level performance on specific tasks \cite{Brown2020} have not yet been implemented or explored for visualisation generation directly from NL. LLMs have revolutionised the field of NL understanding and generation. These models are based on the transformer architecture \cite{vaswani2017attention}, which has demonstrated remarkable successes in tasks such as sentiment analysis, question-answering, and language generation owing both to the effectiveness of the architecture, but also due to them being trained on vast amounts of data. The data used to train these models typically comes from the internet and can include websites, books, and code repositories. As a result of being trained on such a large amount of data, LLMs have developed a comprehensive understanding of the structure and meaning of language, allowing them to perform tasks in a highly sophisticated manner, but also importantly to this study, the ability to generate code in response to NL requests. For these reasons, LLMs offer the potential to accurately understand free-form NL input and the capability to convert it into functional code that generates suitable visualisations which correctly pair with the underlying data types.

\subsection*{Contribution}
\label{sec:contribution}

This work is the first of its kind to propose an end-to-end NL2VIS solution which converts free-form conversational language into visualisations via LLMs. 
The present work leverages the most recent advances in LLM technologies and AI in general, specifically investigating ChatGPT and GPT-3, which are considered state-of-the-art \cite{Brown2020}. The advantage of using pre-trained LLMs for this task is that they offer not only accuracy gains in robustly understanding user requests, even when malformed, but they also result in accelerated development turnarounds and a significant decrease in costs.
Despite the capabilities of these models to display human-like performance on specific tasks, the integration of LLMs with NLIs for visualisation has not been explored in published literature.
Therefore, this study seeks to examine the capability and comparative performances of two types of GPT-3 models and ChatGPT for NL2VIS tasks that also includes the automatic selection of chart types, through numerous experiments and examples. 
Our experiments demonstrate the potential of the LLMs to enhance the performance of NL2VIS in terms of accuracy, efficiency, and cost reduction, thus potentially minimising the need to devise new language models for this problem going forward. Our work also demonstrates how LLMs can be used in a manner that is data-privacy preserving and security-aware, making the approach generalisable to all types of datasets, irrespective of confidentiality concerns. In the process, we show how prompts for LLMs can be engineered to elicit desired outputs. Furthermore, the system developed in this study has also been made publicly available through an online application for testing and experimentation, with the ability for users to upload their datasets and generate visualisations\footnote{ Chat2VIS is currently hosted on Streamlit Cloud and can be accessed via https://chat2vis.streamlit.app/}.

\begin{figure}[h!]
\centering
\includegraphics[scale=0.5]
    {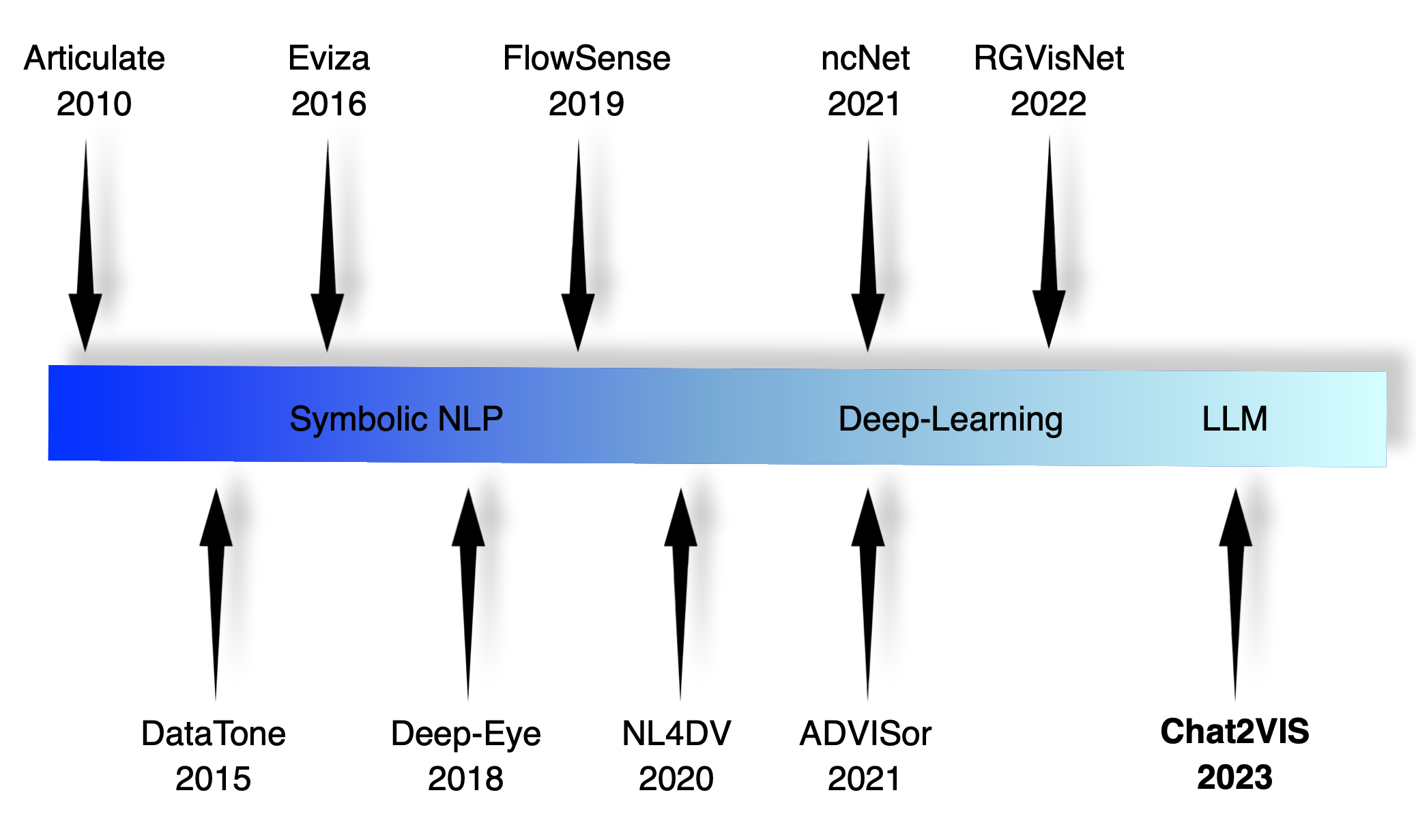} 
 \caption   {Timeline depicting the recent evolution of NL2VIS systems and the proposed Chat2VIS system.}
    \label{fig:timeline}
\end{figure}

\section{Related Work}

In recent years, the idea of using NL as a way to create visualisations has gained significant attention within the field of data visualisation \cite{voigt2021challenges,luo2021natural}. The use of NLIs has also grown in popularity in commercial software as a means of improving the usability of visualisation systems. Various tools, such as IBM Watson Analytics, Microsoft Power BI, Tableau, ThoughtSpot, and Google Spreadsheet, have to varying degrees implemented NLIs that allow users to generate visualisations in response to NL queries or prompts \cite{narechania2020nl4dv,luo2021natural,wang2022towards,nvBench_SIGMOD21,tang2022sevi} indicating the level of demand for this innovation. These are early commercial iterations of this technology as these systems typically constrain NL interactions to data queries and standard chart types, and do not support more complex or open-ended visualisation tasks  \cite{wang2022towards}.

\begin{figure}[h!]
\centering
\includegraphics[scale=0.9]
    {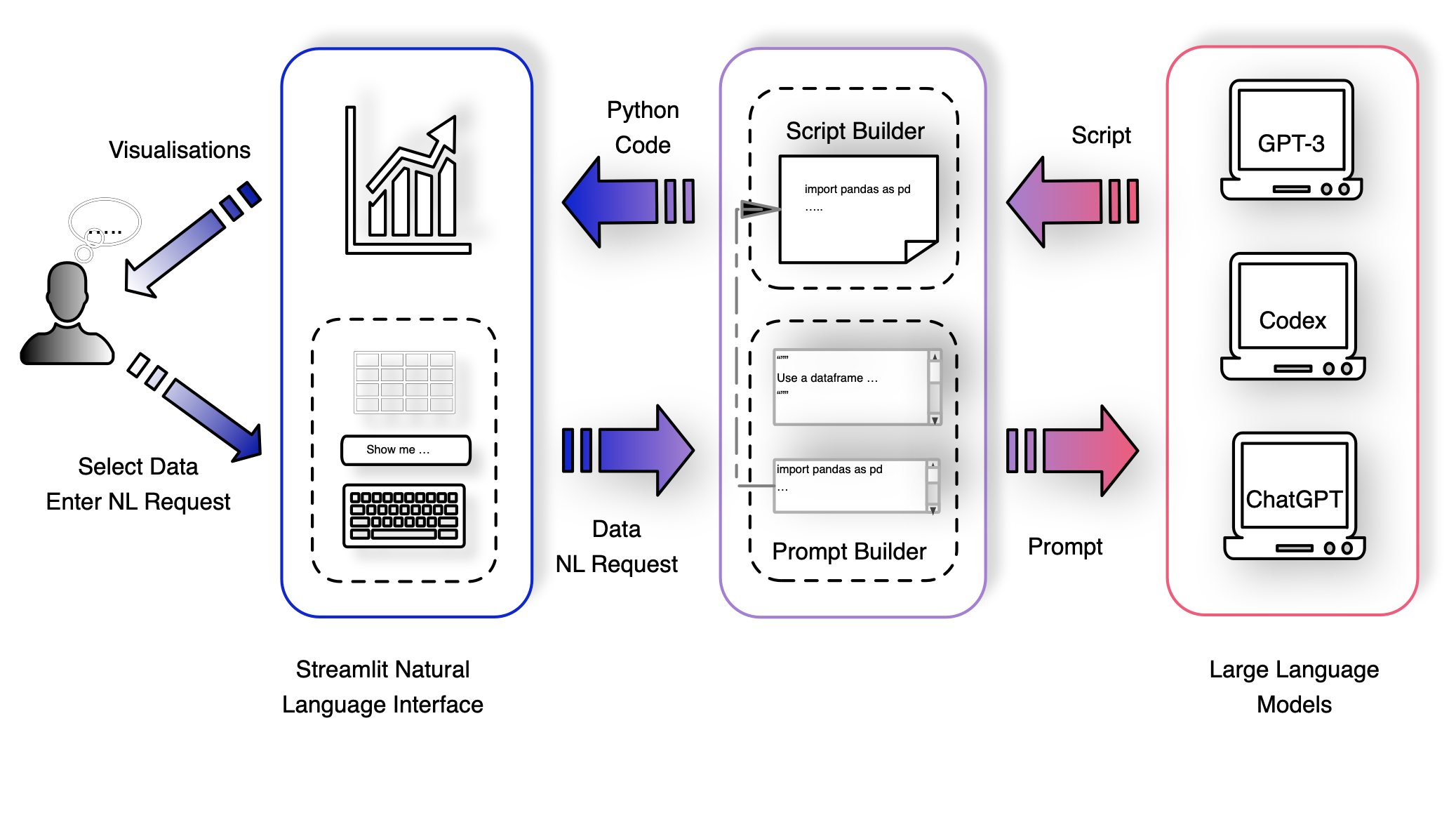} 
 \caption   {The Chat2VIS architecture 
    converting free-form query text into visualisations using the large language models GPT-3, Codex, and ChatGPT.}
    \label{fig:overview}
\end{figure}

NL modelling techniques, which underpin NL2VIS,  
can broadly be categorised into traditional symbolic-based NLP approaches which rely on explicit rules and representation of the language structure and the emerging neural machine translation methods which rely on language models developed typically through deep learning \cite{nvBench_SIGMOD21}.  Fig. \ref{fig:overview} depicts the evolution of NL2VIS systems recently, and their trend towards more sophisticated machine learning approaches, where we posit that the logical trajectory is pointing in the direction of using the most advanced AI systems like LLMs for language understanding and code generation.

\subsubsection*{Symbolic NLP approaches}

Symbolic approaches to NLP can involve heuristic, rule-based and probabilistic grammar-based approaches for parsing and understanding NL. Heuristic algorithms use pre-defined rules or heuristics to find approximate solutions for complex problems like NL and tend to be less accurate than other methods \cite{liu2020extracting}. Meanwhile, rule-based approaches rely on predefined rules created by experts for understanding NL. They are more accurate and reliable but, also less flexible in handling complex queries \cite{voigt2021challenges}. Probabilistic grammar-based approaches use formal grammar rules and probability distributions over the possible parses of a given input. These approaches are considered more accurate as well as flexible than the previous approaches but require more computational resources. Each of these approaches presents complexities and can be time-consuming and challenging to resolve especially for developers without prior experience  \cite{narechania2020nl4dv}. 

The majority of the earlier systems have been developed using these approaches \cite{tang2022sevi}, including Articulate \cite{sun2010articulate}, DataTone \cite{gao2015datatone}, Eviza \cite{setlur2016eviza}, and Deep-Eye \cite{qin2018deepeye}, with each using distinct methodologies for mapping NL queries to visualisations. Certain systems also allow for user interaction to manage ambiguities.  
Meanwhile, recent state-of-the-art studies like NL4DV \cite{narechania2020nl4dv} and FlowSense \cite{yu2019flowsense} using symbolic NLP approaches have relied on semantic parsers such as NLTK \cite{loper2002nltk}, NER, and Stanford CoreNLP \cite{manning2014stanford}, to automatically add layers of valuable semantic information, such as parts-of-speech (PoS) tagging and named entity recognition to the NL input which improves accuracy.
NL4DV \cite{narechania2020nl4dv}, an open-source Python toolkit, accepts a NL queries for a given dataset and outputs a JSON object with keys for dataset attributes, analytical tasks, and Vega-Lite visualisation specifications. This framework enables those who lack NLP expertise to harness NL2VIS concepts by creating NLIs or rendering the output in their own systems.

\subsubsection*{Deep-learning model approaches}

A more promising route has been highlighted by recent advancements in NL2VIS systems which have shifted the focus towards neural end-to-end models that leverage deep learning. These approaches combine the processes of language understanding, reasoning and chart generation in a single system and are closer to the proposed system in this work. These approaches aim to achieve greater robustness, flexibility and adaptability compared to traditional methods \cite{voigt2021challenges}.

One notable system developed along these lines is ADVISor \cite{liu2021advisor}. ADVISor comprises multiple customised deep learning modules for determining the necessary visualisation data and appropriate attributes, as well as the filter and aggregation operations. These models are underpinned by the large language transformer-based BERT \cite{devlin2018bert} model which performed the vectorisation of user input and the data attribute names for subsequent steps. Meanwhile, the chart type is chosen based on a predefined rule-map. 

ncNet \cite{luo2021natural} is a novel approach which also uses transformer-based models and visualisation-aware optimisations. ncNet is a machine learning model that is trained using the nvBench \cite{luo2021nvbench} dataset, which maps natural language queries to visualisations. The system accepts the NL query and an optional chart template as an additional input to constrain the possible visualisations being outputted. The approach has been evaluated through quantitative evaluation and user study and has shown promising accuracy in the nvBench benchmark. Recently, this system has been extended to include speech-to-visualisation capabilities \cite{tang2022sevi}.

Meanwhile, a system called RGVisNet \cite{song2022rgvisnet} was developed that decouples the NL2VIS process into two subtasks which consist of a hybrid retrieval and a generation framework. The first part of the system retrieves the most relevant visualisation query from a large-scale visualisation codebase which serves as a candidate prototype which is refined in the next step by a GNN-based deep-learning model.

\subsection*{Summary of Literature and Research Aims}
The trends in NL2VIS research highlight that focus is moving towards using transformer-based deep learning models and end-to-end solutions, and away from the complex task of engineering symbolic-based language models. Recent advancements, such as the use of pre-trained language models like BERT and domain-specific models like ncNet, have shown promising results in various benchmarks. However, there is a gap in the literature exploring the recent state-of-the-art pre-trained LLMs which are significantly larger and more sophisticated. This work aims to address this gap and examine the potential to simplify the NL2VIS pipeline, while making it more robust for free-form NL and complex visualisation tasks. Additionally, while advanced systems like ADVISor \cite{liu2021advisor} and  ncNet \cite{luo2021natural} rely on various explicit mechanisms for determining which types of visualisations are to be rendered, the proposed system addresses this gap by investigating the ability to delegate the chart selection decision-making to the AI component.

\subsubsection*{Research questions}

In light of the existing literature, this study poses the following research questions:

\begin{itemize}
    \item (RQ1) Do current LLMs support accurate end-to-end generation of visualisations from NL? 
    \item (RQ2) How can LLMs be effectively leveraged in order to elicit the generation of correct and appropriately rendered charts? 
    \item (RQ3) Which LLMs tend to perform more robustly to NL prompts? How do they perform against other state-of-the-art approaches?
    \item (RQ4) What are the limitations of using LLMs for NL2VIS and future research directions?
\end{itemize}

\section{Methodology}
\label{sec:methodology}

In this study, we explored the ability of three OpenAI LLMs to generate Python scripts for visualising data based on NL queries without explicit direction as to which types of graphs to generate. The models chosen for this investigation include the most advanced model family, Davinci, specifically the GPT-3 model \textit{"text-davinci-003"} and the Codex model \textit{"code-davinci-002"}, as well as the most recent addition, ChatGPT\footnote{OpenAI documentation refers to "text-davinci-003", "code-davinci-002" and ChatGPT models as belonging to the GPT-3.5 series https://platform.openai.com/docs/model-index-for-researchers }.

The Davinci model family consists of billions of parameters\footnote{The reported number of parameters is 175B}, and is widely considered to be the most capable of all available models in its ability to follow instructions. This model family is based on GPT-3, with the Codex model receiving additional training data from a massive quantity of GitHub repository code. This makes the Codex model particularly well-suited for translating NL into code, being proficient in over a dozen programming languages, with Python being the target language in this study.

Fig. \ref{fig:overview} depicts the overview of the developed Chat2VIS system. A user enters a NL query via a Streamlit NLI app which is an open-source Python framework for web-based dashboards. The query is combined with a prompt script which engineers a suitable prompt for a selected dataset. The prompt is forwarded to selected LLMs, which return a Python script that is subsequently rendered within the Streamlit NLI.

\subsection{Natural Language Interface}

\begin{figure}
\centering
\includegraphics[scale=0.9]
{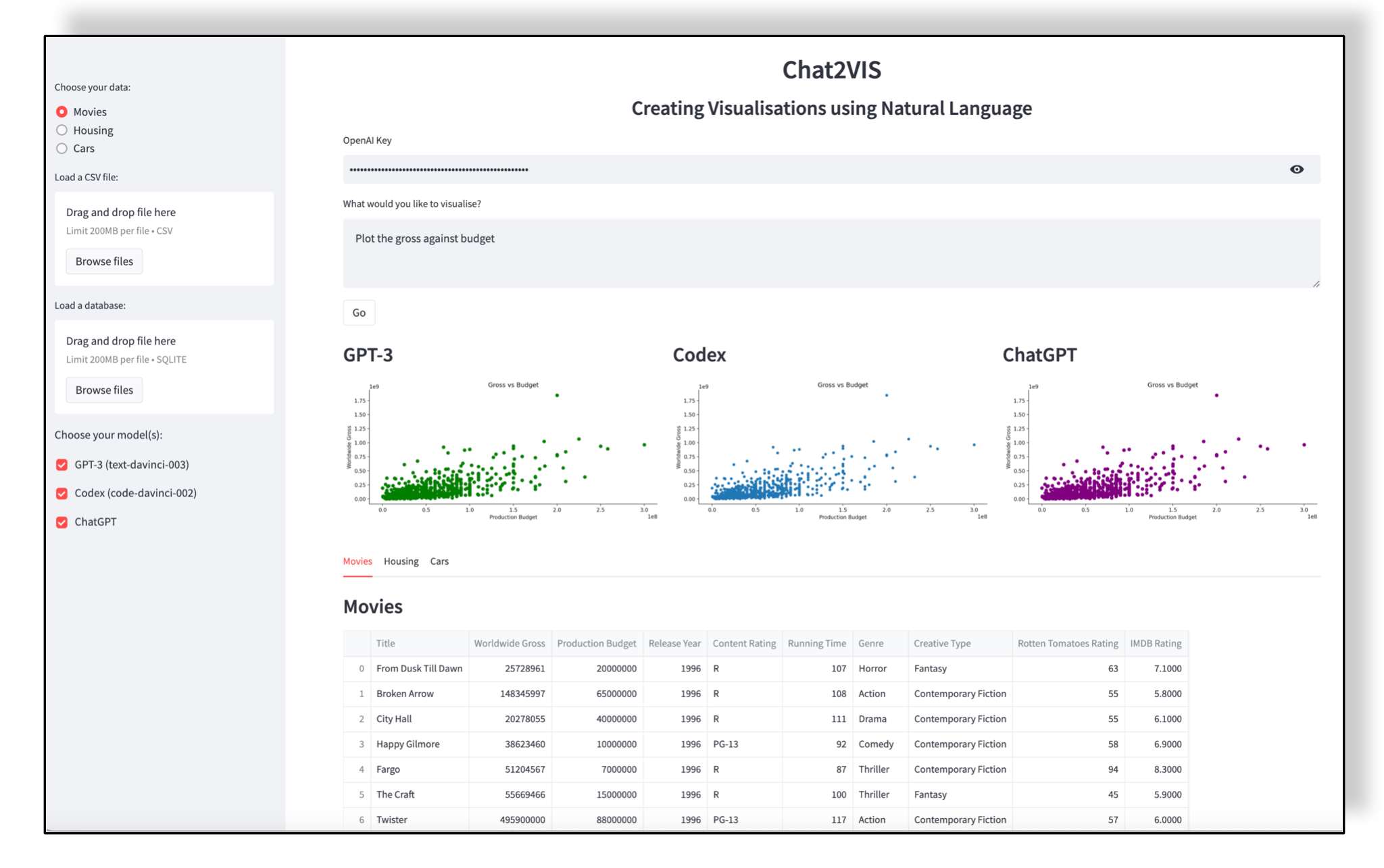} 
\caption{The Streamlit Chat2VIS Natural Language Interface enabling dataset visualisations from free-form text queries.}
\label{fig:interface}

\end{figure}

The interface for the Chat2VIS software artefact used in this study is depicted in Fig. \ref{fig:interface}. The interface enables users to select a dataset and enter free-form text describing their data visualisation intent. The side toolbar provides the functionality to import additional CSV files and SQLite databases, with options to choose the desired LLMs\footnote{At the time of writing this manuscript, no official  API for ChatGPT was available and the process of handling the request for this LLM using Chat2VIS was executed manually through the OpenAI ChatGPT interface.  Use of ChatGPT within Chat2VIS hosted on Streamlit cloud is temporarily suspended until the immanent public release of the official OpenAI API.} to render the visualisations.
An OpenAI Access Key is required to access the models and must be entered prior to querying.  
An input box is provided for entering the NL free-format text. Visualisations are presented for each selected model, with the actual dataset also shown to the users.

\subsection{Prompt Engineering}

The most effective method for obtaining the desired output from the LLM is to use the "show-and-tell" technique\footnote{https://beta.openai.com/docs/guides/completion/prompt-design} by supplying within the prompt an example together with instructions. The proposed system generates a LLM prompt consisting of two parts: (1) a Description Prompt built from a Python docstring and declared using triple double quotes """ at the beginning and the end of the definition, (2) a Code Prompt comprising of Python code statements that provide guidance and a starting point for the script.

\begin{figure}
\centering
 \includegraphics[scale=0.5]
 {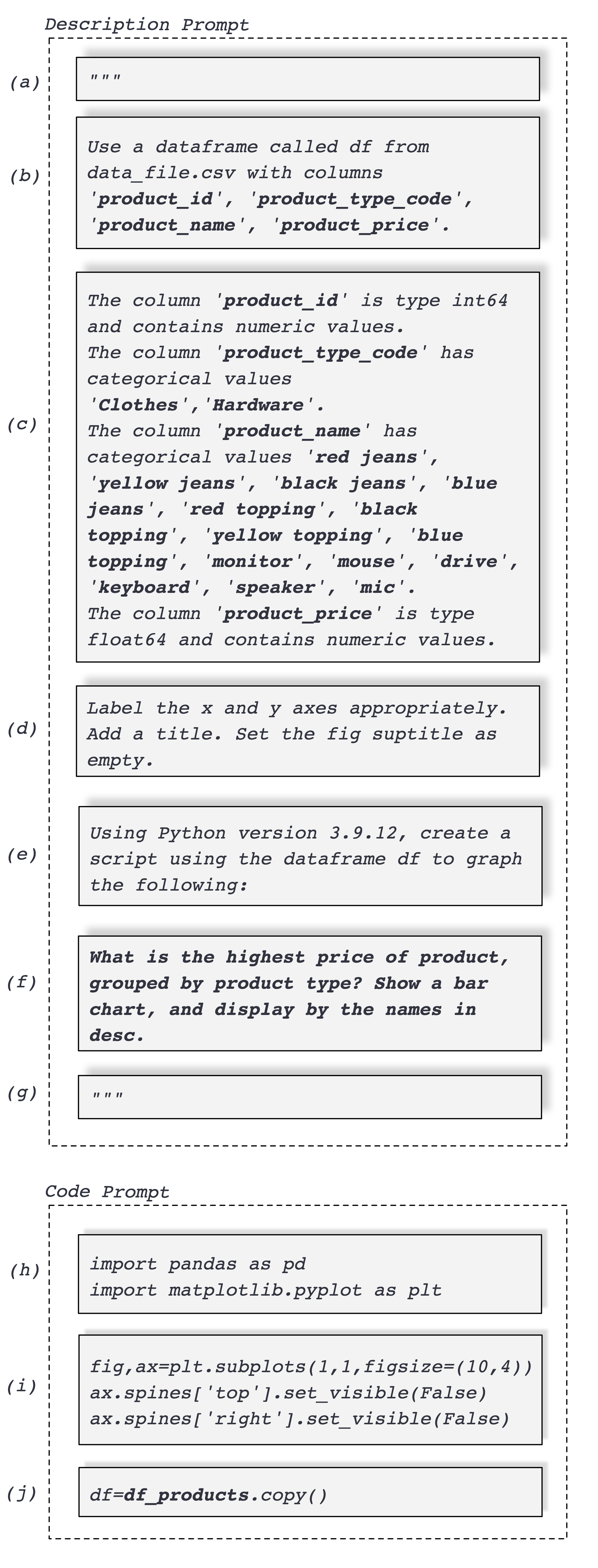} 
 \caption{Description and Code Prompts built from combining the free-form text query with the selected dataset to be submitted to the LLMs.}
    \label{fig:primer}
\end{figure}

The structure of the prompt is presented in Fig. \ref{fig:primer} and is  described by way of an example
dataset, using the \textit{products} table from the nvBench 
database\footnote{https://github.com/TsinghuaDatabaseGroup/nvBench/blob/main/nvBench\_VegaLite/VIS\_6.html}
\textit{department\_store}, pre-loaded into dataframe \textit{df\_products}. For context, the example dataset lists the prices for a selection of clothing products such as coloured jeans and tops, together with hardware products like monitors and keyboards. There are four columns, \textit{product\_id}, \textit{product\_type\_code}, \textit{product\_name} and \textit{product\_price} which have data types \textit{int64}, \textit{object}, \textit{object} and \textit{float64} respectively.

The Description Prompt and the Code Prompt shown in Fig. \ref{fig:primer} are depicted in components in order to aid their description. The bolded type highlights substitution values which are variable and dependent on the chosen dataset and is thus specific to our illustration. These provide an overview of the dataframe to the LLM, listing column names, their data types and categorical values, which assists the LLM in understanding the context, while maintaining data privacy by withholding the actual raw values. Each component of the proposed prompt is described as follows:

\begin{enumerate}
    \item The Description Prompt is initiated with a Python docstring Fig. \ref{fig:primer}(a)
    
\item In Fig. \ref{fig:primer}(b) we explicitly inform the LLM to use a dataframe (a tablular data object) with the name \textit{df} which enables us to refer to this dataframe by a specific name, thereby avoiding any confusion that might arise if the LLM were to assign a different name to the dataframe. Even though we opted to utilise a pre-loaded CSV file, there may be instances when the LLM (typically ChatGPT) includes code for loading the file. By anticipating the file name as data\_file.csv, we can easily identify and eliminate this code if required before execution.

\item The Description Prompt in Fig. \ref{fig:primer}(c) consists of one entry for each column indicating its data type. If a column with an object data type has less than 20 distinct values, it is deemed as a categorical type, and its values are enumerated in the prompt. This listing could aid the LLM in identifying keywords for certain requests, for instance, prompts relating to keyboards or black jeans in our example.

\item In Fig. \ref{fig:primer}(d) we ask the LLM to decide on appropriate naming for the x and y 
axes, and the plot title. In some cases, when working with grouped data, particularly box plots, the LLMs add a plot super-title unnecessarily. To address this, we recommend removing it by using positive instructions and explicitly setting it to empty.

\item To encourage correct syntax, we include the Python version as shown in Fig.  \ref{fig:primer}(e).

\item The Description Prompt concludes with an instruction to create a plotting script with the supplied user query in Fig. \ref{fig:primer}(f). 

\item For this demonstration, the submitted NL query from nvBench is: \textit{What is the highest price of product, grouped by product type? Show a bar chart, and display by the names in desc.}

\item  The Python docstring in Fig. \ref{fig:primer}(g) closes off the Description Prompt

\item The Code Prompt begins with import statements for the required Python packages we would like the LLMs to use in Fig. \ref{fig:primer}(h).

\item To foster uniformity in the plot layout, we ask for a single subplot with a fixed figure size in Fig. \ref{fig:primer}(i) in an attempt to render equivalently sized plots on the interface.

\item By assigning a copy of the named dataframe in Fig. \ref{fig:primer}(j) to the variable \textit{df} in the Code Prompt, ensures consistency within the script and enables the original dataframe to be retained for further querying.

\end{enumerate}

Once formulated, the two prompt elements are amalgamated, with the resulting string submitted to the LLMs via the text completion endpoint API with settings shown in Table. \ref{table:APIParameters}.  

\begin{table}[h!]
\centering
\caption{API Parameters.}   
\begin{tabular}  
{p{60pt}p{200pt}}
\hline 
Parameter & Setting \\
\hline
engine & text-davinci-003 or code-davinci-002 \\
prompt& <constructed NL prompt string>\\
temperature&0\\
max\_tokens&500\\
stop&plt.show() \\
 \hline\end{tabular}
\label{table:APIParameters}
\end{table}

Unspecified parameters remain at their default values.  With many variations to Python scripting, we set the model temperature to 0 to help ensure the LLM is less creative and more consistent in its code generation.  To encourage the model to avoid any unnecessarily long script responses, a token limit of 500 is enforced for responses.  We believe this limit is sufficient for generating a Python script within the context of this study. A stopping point is specified to avoid returning multiple script examples and code alternatives.

\subsection{Script Refinement and Rendering}

On the return of the script from the API for each model, the Code Prompt is inserted at the start and the Python code may be edited to eliminate unnecessary instructions. It is rendered on the interface for each LLM.  To further enhance the visualisations, users can refine their NL query, including requesting alternative chart types, plot colours, labels etc.

\subsection{Chat2VIS Evaluation}

The capabilities of the proposed Chat2VIS system to render visualisations based on NL input via LLMs, and their unique decision-making skills in autonomously selecting appropriate charting elements is demonstrated over six case studies, covering five datasets. 
Four of these case studies are reproduced from examples in existing literature, comparing our visualisations with those from prior studies using nvBench SQLite databases\footnote{https://sites.google.com/view/nvbench} and the NL4DV Python package\footnote{https://nl4dv.github.io/nl4dv/documentation.html}.
One case study is taken from NL4DV \cite{narechania2020nl4dv}, one from ADVISor \cite{liu2021advisor} in which the authors also use NL4DV in their evaluation, and the final two using databases from nvBench \cite{luo2021nvbench}.

In the remaining two case studies, we 
test the capabilities of the LLMs and the prompt scripts to handle misspecification in the form of typographical errors as well as acute underspecification, where the ability of the LLMs to exhibit reasoning and assumptions in the context of the priming text is explored. 

The results are inspected and evaluated visually for correctness and suitability. All case study examples are also reproducible from the online web app. The non-deterministic nature of the LLMs does occasionally lead to variability in plot generation even when an identical prompt is resubmitted.

\section{Results}
\label{sec:results}

\subsection{Case Study 1: Department Store Dataset}
\begin{figure}[h!]
\centering
\includegraphics[scale=0.9]
{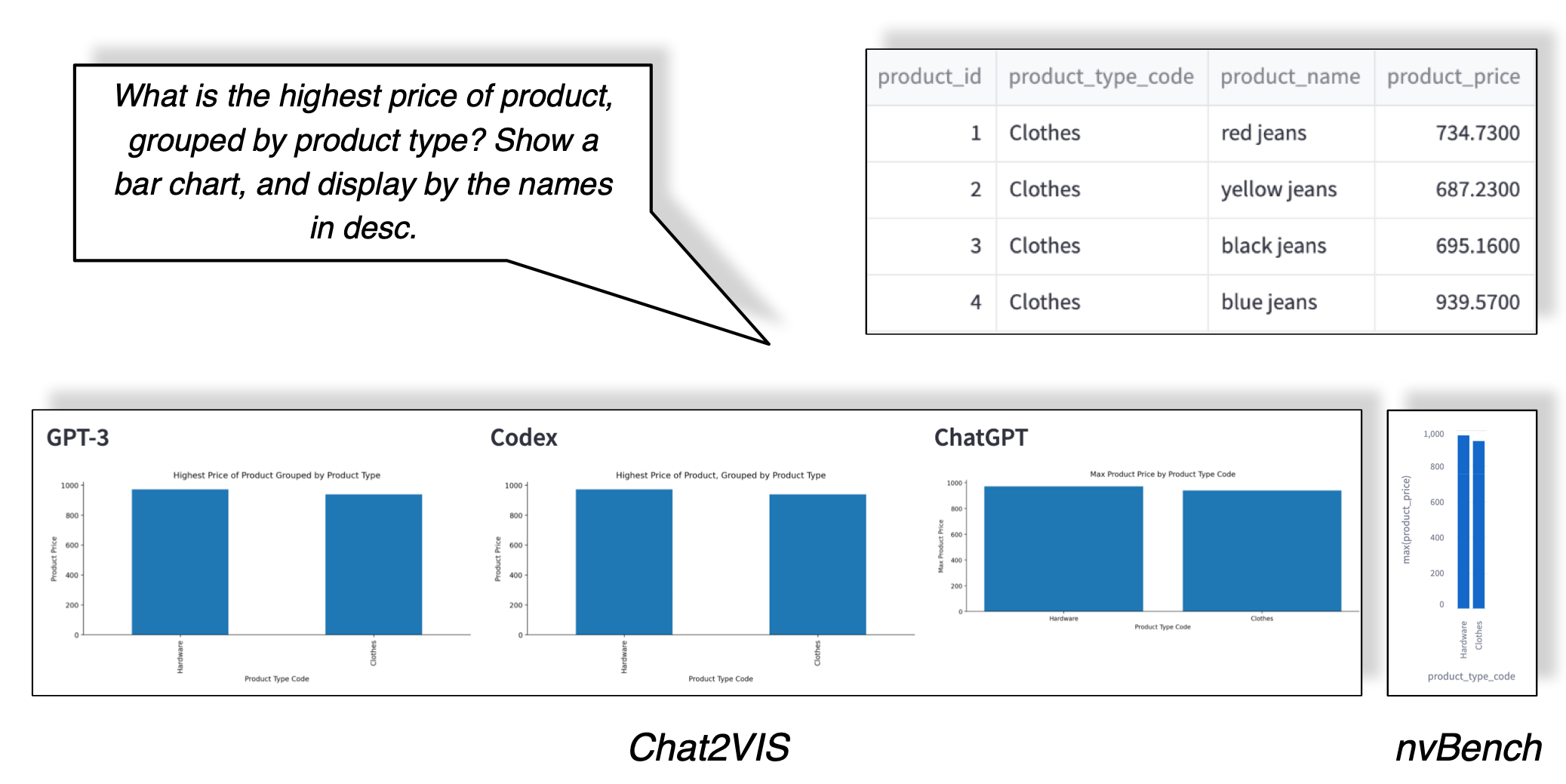} 
\caption{Case Study 1: Department Store Dataset}
\label{fig:eg_department_store}    
\end{figure}

The first example is based on the \textit{products} table illustrated above in the description of the prompt engineering which originated from the nvBench \textit{department\_store} database. The test query is as follows: \textit{"What is the highest price of product, grouped by product type? Show a bar chart, and display by the names in desc."}. The query is categorised by nvBench as an \textit{easy} visualisation for NLI systems. Fig. \ref{fig:eg_department_store} shows results generated by Chat2VIS alongside the correct nvBench visualisation. GPT-3 and Codex produce identical results, with ChatGPT rotating labels for ease of reading and arguably providing a slightly more comprehensive title.  All 3 LLMs provide more informative \textit{x} and \textit{y} axis labelling and titles than the ground truth example from nvBench.

\subsection{Case Study 2: Colleges Dataset}
This dataset holds information on students and staff at U.S. public and private colleges. The \textit{colleges} dataset\footnote{https://github.com/nl4dv/nl4dv/blob/master/examples/assets/data/colleges.csv} was used by \cite{narechania2020nl4dv} to demonstrate the Vega-Lite editor for rendering queries visualised by NL4DV.
The output from their toolset rendered via Streamlit
is compared with Chat2VIS results. 
The broadness of the submitted query used in their study \textit{"Show debt and earnings for Public and Private colleges." } allows flexibility for different interpretations of the request. 

\begin{figure}[h!]
\centering
\includegraphics[scale=0.9]
{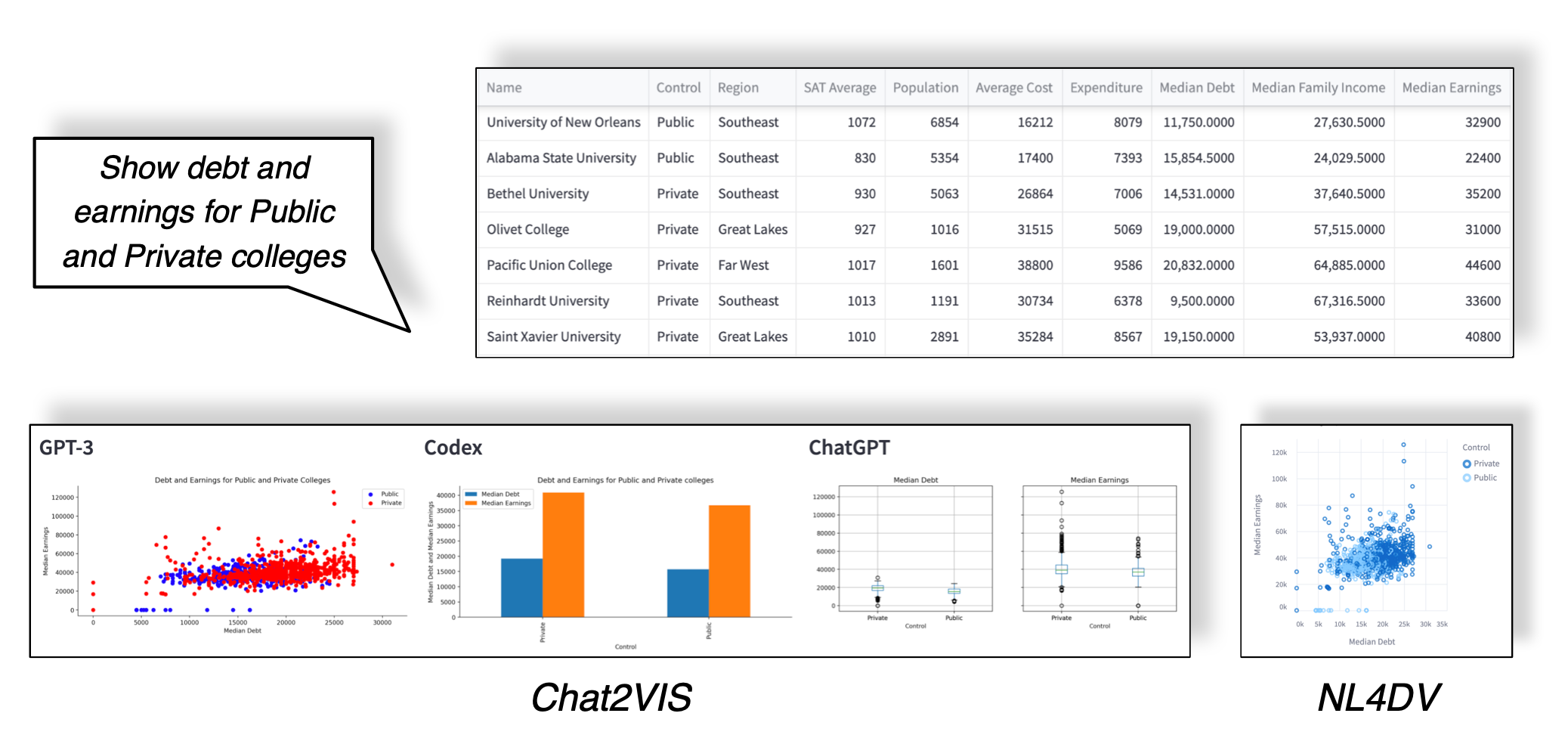} 
\caption{Case Study 2: Colleges Dataset}
\label{fig:colleges}
\end{figure}

The figure (Fig. \ref{fig:colleges}) demonstrates that GPT-3 generated a scatter plot, similar to 
that produced by NL4DV, to represent the relationship between median debt and median earnings for all colleges by type. Codex interpreted the query by creating a bar chart to show the mean value of debt and earnings for each college type. ChatGPT focused on the distributions of debt and earnings by creating a box-and-whisker plot for public and private colleges. The models were able to differentiate between public and private colleges using the "Control" column, which was not easily recognisable as a college type. However by providing categorical values in the prompt, the models were able to accurately categorise the data based on the "public/private" indicator in the column. In addition, the models were required to perform further reasoning in order to arrive at the decision to extract data from the "Median Debt" and "Median Earnings" columns,
having only been queried to show debt and earnings, thus illustrating the capability of AI inference potentials.

\subsection{Case Study 3: Energy Production Dataset}

The next set of visualisations are depicted on the Energy Production dataset described in the ADVISor  study and compared to those of both ADVISor and 
 NL4DV outputs.  The dataset details coal, oil, gas, and nuclear energy production in megawatt-hours per person per year from 2000 to 2011 for an unspecified country, including population statistics. 

\begin{figure}[h!]
\centering
\includegraphics[scale=0.9]
    {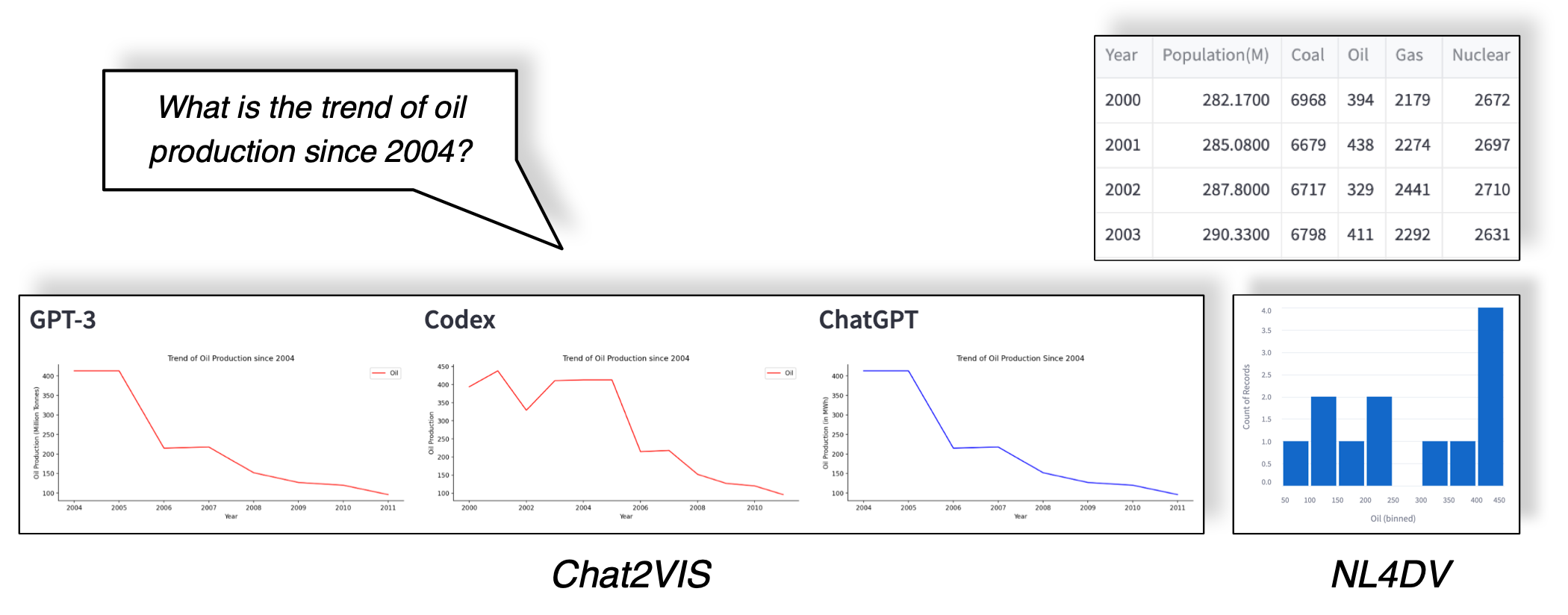}
    \caption{Case Study 3: Energy Production Dataset}
    \label{fig:oil}
\end{figure}

We use here the test query from the ADVISor study, \textit{"What is the trend of oil production since 2004?"}. The results from Chat2VIS are shown in Fig. \ref{fig:oil} together with the output from
NL4DV rendered via Streamlit. The results can be 
 compared with the ADVISor plot presented in their study.  
All three LLMs select a line plot as the most suitable style of plot for this query, with GPT-3 and ChatGPT correctly showing data from 2004 onward. Codex, however, neglects to incorporate this detail into its code generation and has depicted data from 2000 onwards.  
NL4DV has produced an incorrect visualisation due to its semantic parsing limitations which lacks flexibility, as mentioned in \cite{liu2021advisor}. 
The ADVISor plot also selected the correct chart type, but like Codex, it was unable to filter the data to include 2004 onwards. It did, however, highlight the data points from 2004 onwards drawing attention to the oil trend and the selected data range.

\subsection{Case Study 4: Customers and Products Contacts Dataset} 

The following example provides a demonstration of a more complex NL2VIS task from the \textit{products} table in the \textit{customers\_and\_products\_contacts} database\footnote{https://github.com/TsinghuaDatabaseGroup/nvBench/blob/main/
nvBench\_VegaLite/VIS\_299.html}, 
which the nvBench 
example
benchmark classifies as \textit{Extra Hard}. The query is  \textit{"Show the number of products with price higher than 1000 or lower than 500 for each product name in a bar chart, and could you rank y-axis in descending order?"}. 

\begin{figure}[h!]
\centering
\includegraphics[scale=0.9]
    {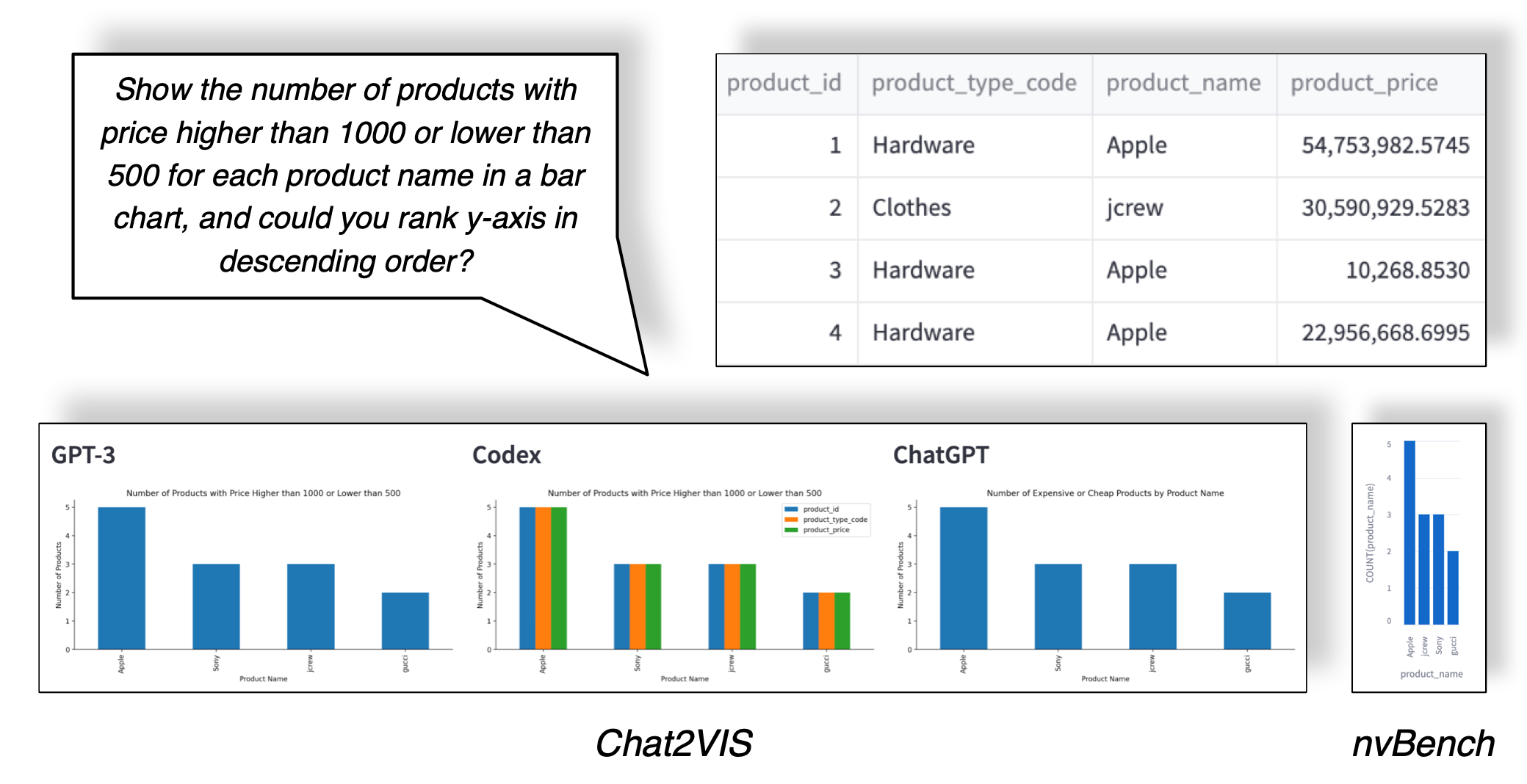} 
    \caption{Case Study 4: Customers and Products Contacts Dataset}
    \label{fig:customers_products_contacts}
\end{figure}

Fig. \ref{fig:customers_products_contacts} shows all three models generate similar visualisations to the ground truth example specified in nvBench. ChatGPT provides a slightly more informative title, conveying that products with a price higher than 1000 are "expensive" and those lower than 500 are "cheap".
Despite Sony and \& jcrew products swapped in comparison to nvBench, both have a value of 3 and are plotted accurately.
The visualisation from Codex, while correct, is sub-optimal, requiring some further improvement in its coding structure to eliminate the multiple bar plotting. All three plots show more informative axis labels than their nvBench counterpart. The example demonstrates the high capability levels of Chat2VIS to handle challenging NL queries.

\subsection{Case Study 5: Misspecified Prompts}

The following example illustrates the robustness of Chat2VIS to input errors that take the form of typographical mistakes. 
The movies\footnote{https://github.com/nl4dv/nl4dv/blob/master/examples/assets/data/movies-w-year.csv} dataset is used here
and contains information on movies released between 1996 and 2010, including details such as financials, ratings, and classifications. The ideal query in this example is  \textit{"Plot the number of movies by genre"} however, the system is prompted with \textit{"draw the numbr of movie by gener"}.
 
\begin{figure}[h!]
\centering
\includegraphics[scale=0.9]
    {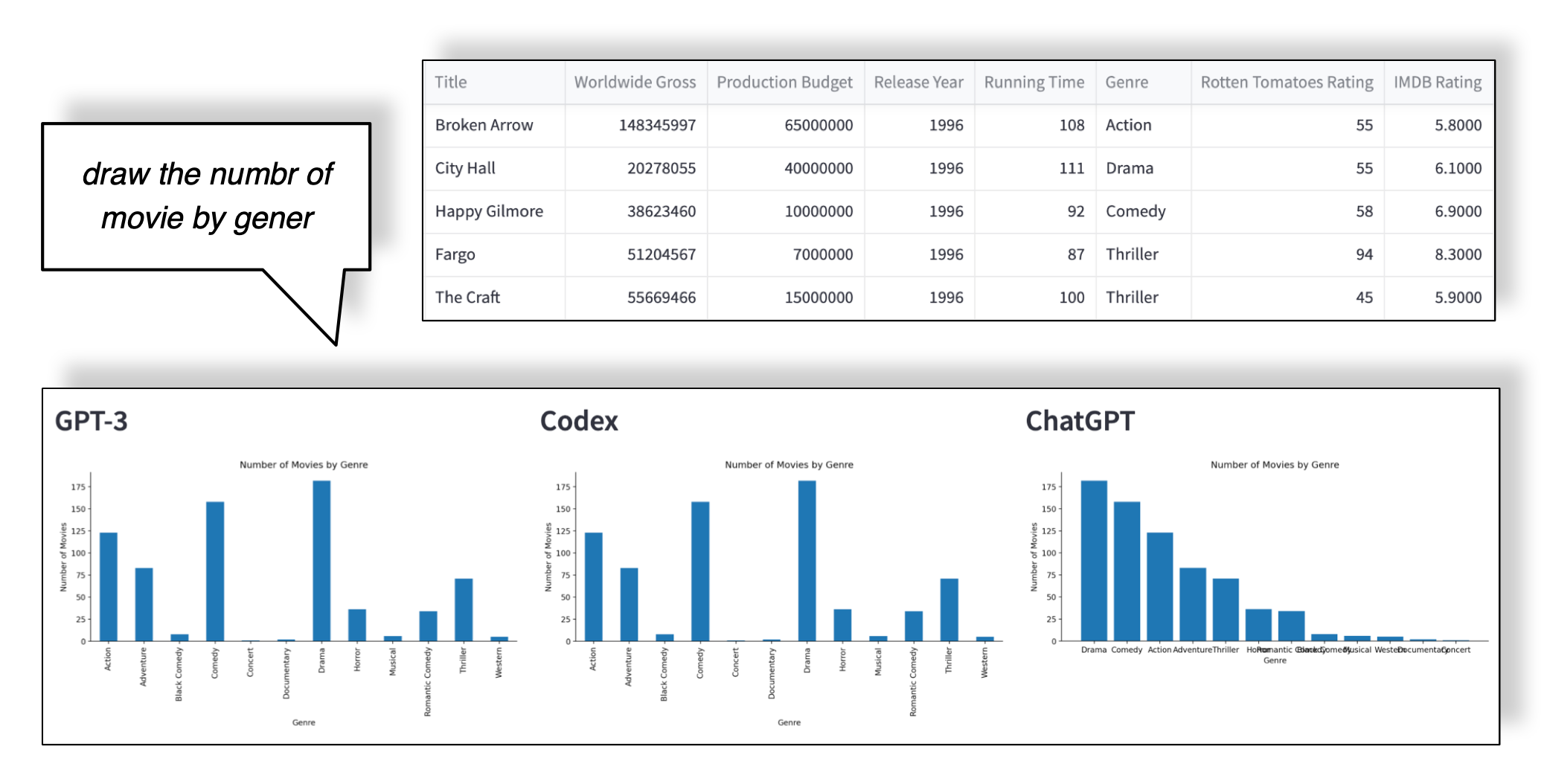} 
    \caption{Case Study 5: Misspelled Prompts}
    \label{fig:spelling}
\end{figure}

The results are shown in Fig.\ref{fig:spelling}. The correct results across all three LLMs highlight the ability of the LLMs to interpret language even in the presence of multiple typographical errors and misspecification, thus emphasising their robustness.
However, from the point of view of clarity of insights, the figure generated by ChatGPT is superior to those of the other models since it has decided to render the results as a rank-ordered bar graph in a descending order, while GPT-3 and Codex presented movies alphabetically.

\subsection{Case Study 6: Underspecified and Ambiguous Prompts}

The movies dataset is again used in this example in order to demonstrate the inference capabilities of the LLMs together with the underlying priming prompts developed for Chat2VIS, to creatively make decisions based on extremely underspecified or ambiguous queries. 

\begin{figure}[h!]
\centering
\includegraphics[scale=0.9]
    {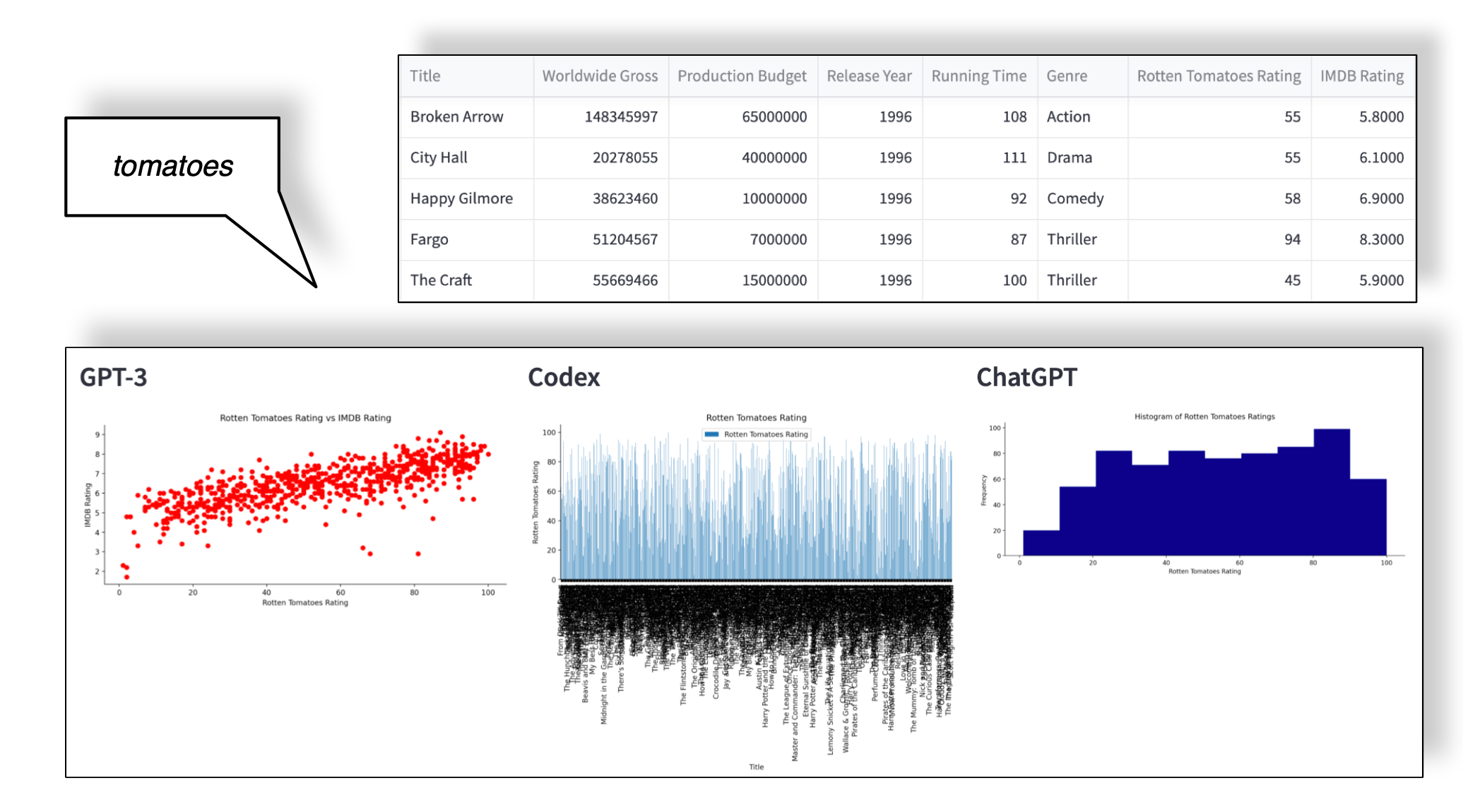} 
    \caption{Case Study 6: Underspecified and Ambiguous Prompts}
    \label{fig:Underspecified}
  \end{figure}

The test query used here is: \textit{"tomatoes"}, which has an association with an existing column in the dataset called \textit{Rotten Tomatoes Rating}. Fig. \ref{fig:Underspecified} demonstrates the results. Remarkably, the figures demonstrate that each LLM, was able to make inferences and produce a figure that connected the results with the \textit{Rotten Tomatoes Rating} despite a lack of direction.  

GPT-3 plots the rating against the IMDb Rating column. It is uncertain whether this column is selected due to it being a rating column or simply because it is the next column in the dataset.  Codex plots the rating for every title, producing an aesthetically unusable visualisation due to overcrowding, while  ChatGPT produces a meaningful distribution plot of the ratings.


\section{Discussion}

The experimental results from the 6 case studies in this work confirm that LLMs can effectively support the end-to-end generation of visualisations from NL when supported by additional prompts, and are therefore, a viable solution for the NL2VIS problem (RQ1). The proposed method exhibits a number of advantages over existing NL2VIS systems which rely on symbolic NLP and deep learning approaches. The primary ones are efficiency, cost savings and accuracy. LLMs offer a pre-trained and simplified solution to the most difficult problem of understanding NL, while also encapsulating the code generation component as well. As such, this approach removes the need to create hand-crafted symbolic NLP solutions, thus saving resources. The abilities of current LLMs raise questions about the necessity to further explore symbolic NLP approaches for providing solutions to NL2VIS as a whole. Meanwhile, the proposed method demonstrated comparable if not indications of superior performances with existing methods, while showing robustness to misspecified and underspecified NL requests. The accuracy of LLMs will only continue to improve with time and therefore they represent the most viable solution for developing NL2VIS system going forward.

To leverage this technology (RQ2), this study demonstrated the importance of prompt engineering in providing the LLMs with clear and concise NL requests. The study demonstrated how effective prompts can be engineered using Description and Code Primer definitions that indicate to the LLMs attributes of the underlying data as well as a coding guide. The results confirm that LLMs can be effectively primed with the proposed prompts to elicit the generation of appropriately rendered charts when accompanying the NL requests.

In terms of performance (RQ3), the preliminary results indicate that the capability of the three LLMs tends not to show large deviations. This is likely due to the fact that the LLMs were trained on similar datasets. 

While the results demonstrate the potential of LLMs for NL2VIS, there are still some challenges to this technology (RQ4), mostly centring around aesthetic features of graphs and variable results. These challenges are addressed individually below.

\subsection{Remaining challenges}
While the proposed framework has performed impressively well and provided a viable solution to the problem of converting free-form NL directly into visualisation, some minor challenges still remain.

\paragraph*{Setting the Plot Background Colour}  
Providing instruction within the engineered prompt to consistently and successfully generate code to change the background colour of a plot proved unsuccessful. With multiple factors at play, ranging from type of plot generated and the Python container it is rendered within, through to software theme settings, no consistent approach was unearthed. 

\paragraph*{Display of Plot Grid Lines} 
The incorporation of grid lines into a plot can enhance its aesthetics and aid in its interpretation. However, generating precise code for the successful rendering of grid lines is often determined by choice of plot type.  As this is not specified within the engineered prompt, it is difficult to inform the LLM of the correct methods and function parameters to render the grid lines successfully.  With thorough experimentation, the LLMs produced varying levels of success when the request for horizontal grid lines was included in the engineered prompt.  Therefore, it is more favourable to request the adjustment of styling elements such as grid lines during the refinement of the NL query, and will be most successful when performed in conjunction with explicitly stating the type of plot to be rendered.

\paragraph*{Specifying Colouring of Plot Lines and Elements} 
Analogously to the problems encountered with rendering grid lines, specifying refinements to other plotting elements such as line colour proved challenging within the engineered prompt. With the dependence of some function and parameter values on plot type selection, LLMs on occasion attempted to invoke unsuitable functions or assign values to unknown parameters while endeavouring to style plot elements as requested.  As with grid line refinement, styling of plot lines will be most successful when combined with plot type and included as an extension to the user query. 

\paragraph*{Variability in Plot Generation} 
The repetition of the same prompt to the language model can result in significant variability in the type of plot generated and its features. This can make it difficult to consistently generate the desired visualisations. With their non-deterministic nature, especially of ChatGPT, is not possible to address this adequately at this stage since parameters that regulate the stochastic processes in its reasoning are not yet publicly available.

\paragraph*{Refining the Prompt for Best Results} 
The generic and verbose nature of the engineered prompt caters to all three LLM models, but experimentation has shown that the ideal prompt for each model can be varied slightly to achieve the best results. Once a specific LLM is selected, the prompt may be optimised and refined for generating the best visualisations. 

\subsubsection*{Study limitations and future work}
The current study included a limited number of case studies consisting of a selection of NL queries and example visualisations. Ideally, a comprehensive evaluation of a system like Chat2VIS would include end-users and a qualitative assessment of the system's usefulness based on their feedback. While the evaluation of NLIs in data visualisation is a complex task in the context of end-user experience, it is the intention of the authors to expand this study and conduct this type of evaluation in the subsequent work. 

Future work will also explore the incorporation of the nvBench benchmark dataset into the refinement of the Chat2VIS capabilities and make use of the dataset for a more comprehensive quantitative analysis of its capabilities across a wider set of queries, thus enabling a more robust comparison against results from prior studies.

\section{Conclusion}

The ability to generate visualisations based on natural language has been a long-standing goal in the field of data visualisation. The development of Natural Language Interfaces has paved the way for advancements in this area making data visualisation more accessible to a broader range of users by allowing them to express their queries and analysis intentions in natural language. However, the process of accurately and reliably translating natural language inputs into visualisations (NL2VIS) has been a challenging problem to solve due to the difficulty in understanding natural language.

This study proposed a novel end-to-end solution for converting free-form natural language into visualisations using state-of-the-art Large Language Models (LLMs). This study explored ChatGPT and its predecessors like GPT-3 and Codex for their ability to solve the task of understanding the queries and both auto-generating code while using their internal inference abilities for selecting the appropriate visualisation types. The proposed system, Chat2VIS, has demonstrated that the use of pre-trained LLMs together with well-engineered prompts, provides an efficient, reliable and accurate solution for the problem of NL2VIS. Chart-type selection is automatic, and the LLMs are able to understand vague user queries as well as those that are malformed. Moreover, the approach is also data-privacy preserving and security-aware, making it generalisable to all types of datasets. 

The present study highlights the viability of LLMs to further the capabilities of existing NLIs for visualisation, providing a simpler pathway towards robust solutions which do not involve the task of defining grammars and customised domain-specific language models for language understanding. The results of this study provide valuable insights for researchers and practitioners in the field of data visualisation and NLIs, and offer simpler and more accurate solutions for making data and insights more accessible to a wider audience.


\end{document}